\documentclass[12pt]{article}
\usepackage{graphicx}
\usepackage[makeroom]{cancel}
\usepackage[noadjust]{cite}
 
\newcommand\pubnumber{SNSN-323-63}
\newcommand\pubdate{\today}

\def\institute{Interuniversity Institute for Higher Energies, Vrije Universiteit Brussel\\
Pleinlaan 2, 1050 Brussels, Belgium}
\def\support{\footnote{Work supported by the Interuniversity Institute for Higher Energies.}}

\def\Title#1{\begin{center} {\Large #1 } \end{center}}
\def\Author#1{\begin{center}{ \sc #1} \end{center}}
\def\Address#1{\begin{center}{ \it #1} \end{center}}

\newcommand\pubblock{\rightline{\begin{tabular}{l} \pubnumber\\
         \pubdate  \end{tabular}}}
\newenvironment{Abstract}{\begin{quotation}  }{\end{quotation}}
\newenvironment{Presented}{\begin{quotation} \begin{center} 
             PRESENTED AT\end{center}\bigskip 
      \begin{center}\begin{large}}{\end{large}\end{center} \end{quotation}}

\def\beq{\begin{equation}}
\def\eeq#1{\label{#1}\end{equation}}
\def\eeqn{\end{equation}}

\def\beqa{\begin{eqnarray}}
\def\eeqa#1{\label{#1}\end{eqnarray}}
\def\eeqan{\end{eqnarray}}

\let\bar=\overbar

\def\Dslash{\not{\hbox{\kern-4pt $D$}}}
\def\dslash{\not{\hbox{\kern-2pt $\del$}}}

\def\msb{{\bar{\ssstyle M \kern -1pt S}}}

\def\tt{t\bar{t}}
\def\SM{tZq-SM}

\def\mT{$m_T^W$}

\begin{document}
\begin{titlepage}
\pubblock

\vfill
\Title{Three lepton signatures from tZq interactions in the SM and top-FCNC at the CMS experiment at $\sqrt{s}=8$ TeV}
\vfill
\Author{ Isis Van Parijs\support\ on behalf of the CMS collaboration}
\Address{\institute}
\vfill
\begin{Abstract}
The search for a top quark in association with a Z boson can be performed in both top pair production as well as single top production. Targeting the three lepton final state, the search can be used to identify the standard model (SM) process of a single top quark associated with a Z boson and to search for flavour changing neutral current (FCNC) interactions. The presented analysis uses a data sample corresponding to an integrated luminosity of $19.7$ fb$^{-1}$ recorded by the CMS experiment at the LHC in proton collisions at $\sqrt{s}=8$ TeV. The cross section for the SM tZq production is measured to be $\sigma(tZq\rightarrow l\nu bl^+ l^- q)= 10^{+8}_{-7}$ fb with a significance of $2.4 \sigma$. Exclusion limits at 95$\%$ confidence level  on the branching ratio of a top quark decaying to a Z boson and an up or charm quark are determined to be $BR(t \rightarrow Zu) < 0.022\%$ and $BR(t \rightarrow Zc) < 0.049\%$ respectively. 
\end{Abstract}
\vfill
\begin{Presented}
$9^{th}$ International Workshop on Top Quark Physics\\
Olomouc, Czech Republic,  September 19--23, 2016
\end{Presented}
\vfill
\end{titlepage}
\def\thefootnote{\fnsymbol{footnote}}
\setcounter{footnote}{0}

\section{Introduction}
When a top quark is produced singly through the t-channel mechanism,  a Z boson can be radiated of one of the quarks from the exchange of W bosons. This leads to a final state consisting of a top quark, a Z boson and an additional quark (\SM). 
The observation of \SM\ production and the subsequent measurement of the production cross section is an important test of the standard model since provides a probe for the coupling of a quark to a Z boson. Moreover, its sensitivity  to the WWZ coupling is comparable to that of the direct WZ boson production~\cite{10}.\\
 In proton collisions at a centre-of-mass energy of $8$ TeV, the total cross section of the $tZq(l\nu bl^+l^-q)$ processes, where $l$ denotes a charged lepton, electron or muon, including
lepton pairs from off-shell Z bosons with invariant mass $m_{l^+l^-} > 50$ GeV, is calculated in the 5 Flavour Scheme to be $\sigma(tl^+ l^- q) = 8.2^{+0.59}_{-0.03}$ (scale) fb~\cite{13}. \\

The flavour changing neutral current of the top quark to the Z boson in the SM is forbidden at tree level and suppressed at higher orders because of the GIM mechanism~\cite{17}. However, by extending the SM, the FCNC branching fraction could be as large as $\mathcal{O}(10^{-4})$, making the search for the production of a top quark in association with a Z boson (tZ-FCNC) a viable hunting ground  for new physics~\cite{21}. 
The most stringent exclusion limit at the 95$\%$ confidence level (CL) on $BR(t \rightarrow Zq)$ is set by the CMS Collaboration and excludes branching ratios above $0.05 \%$~\cite{29}.\\
An effective field theory approach is used~\cite{23}, where only a non zero value for $\kappa_{tZq}$ is being considered.
The interference between single top and $\tt$ FCNC processes increases the total cross section by about 5\% and is neglected~\cite{paper}.

\section{Event selection and Analysis}
For the tZq-SM search, the final state consists of a single top quark, a Z boson and an additional jet preferentially emitted in the forward region of the detector ($|\eta| > 2.4$). The search for tZ-FCNC is performed by combining the single top quark (ST-FCNC) and $\tt$ (tt-FCNC) production modes. This leads to a signature containing a top quark and a Z boson with no extra jets from the matrix element calculation for ST-FCNC, while for tt-FCNC, the FCNC vertex appears in the decay of the top quark, and leads to the same signature as for tZq-SM, but with the non-b jet produced in the central region of the detector. For both searches, the W  and Z boson decays into either electrons or muons, are considered.\\

Signal regions contain events with exactly three isolated leptons ($e$ or $\mu$),  each with a  transverse momentum, $p_T$, greater than 20 GeV and an absolute pseudo-rapidity, $|\eta|$,  smaller than 2.5 ($e$) or 2.4 ($\mu$). There should be two of the same flavour leptons with opposite charge in each event, with an invariant mass  compatible with that of the Z boson ($76< m_{ll} < 106$ GeV). The jets  are required to have $p_T>$ 30 GeV and $|\eta|<$2.4(4.5) for the tZ-FCNC (tZq-SM) search.  For the tZ-FCNC search, two signal regions are defined by requiring exactly one jet that is b tagged (ST-FCNC) or requiring at least two jets from which at least one is b tagged (tt-FCNC). The tZq-SM search requires at least two jets, where at least one jet is b tagged. 
Further reducing backgrounds, the transverse mass of the W boson, \mT\ , is required to be $>$10 GeV, and for the tZ-FCNC search also the missing transverse energy should be $>$40 GeV. 
The background enriched region is defined by selecting events with one or two jets, but vetoing events containing a b tagged jet, increasing the WZ and Drell-Yan  content. \\

\begin{figure}[htb]
\centering
\includegraphics[height=1.5in]{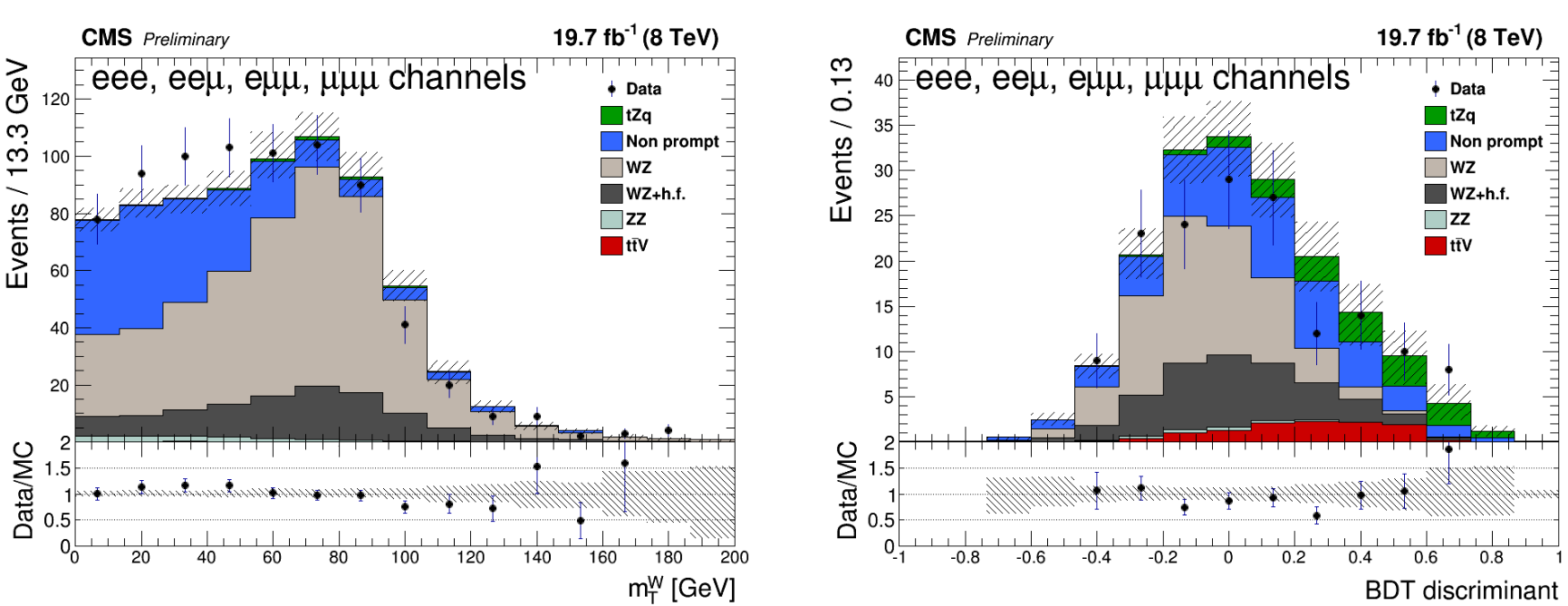}
\includegraphics[height=1.5in]{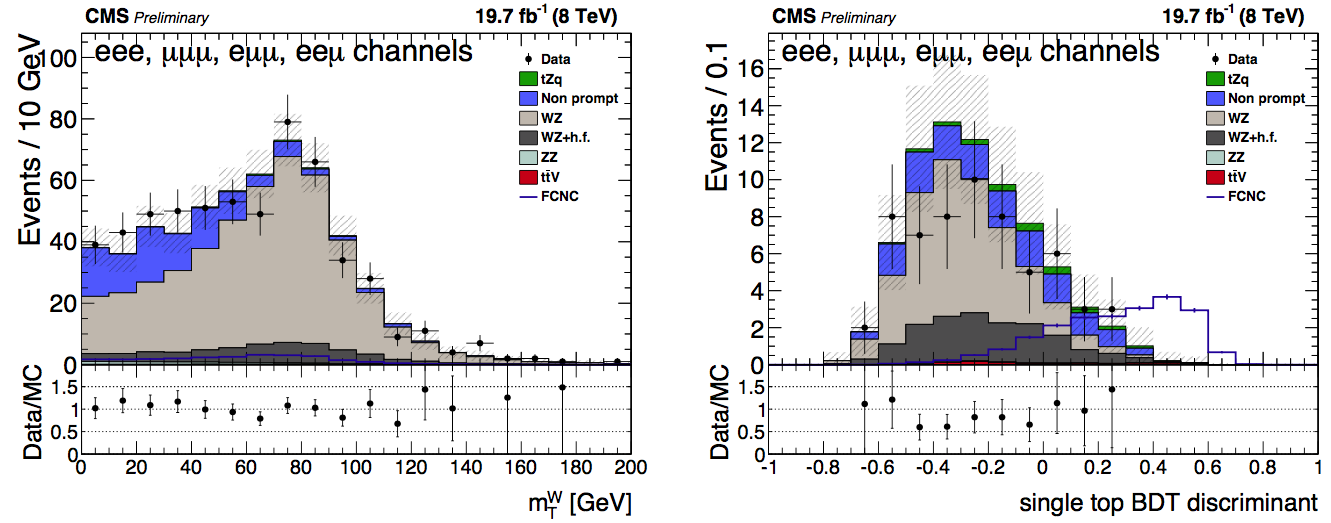}
\includegraphics[height=1.5in]{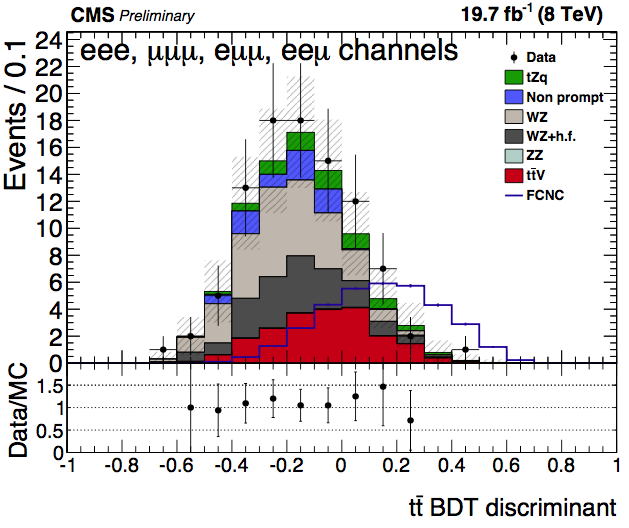}
\caption{Data to prediction comparisons for the tZq-SM (top) and tZ-FCNC (bottom) search after performing the fit for \mT\ distribution in the control region (left), and for the BDT responses in the tZq-SM (right), ST-FCNC (middle), and tt-FCNC (right) signal regions. The four channels are combined. An example of the predicted signal contribution for  $BR(t\rightarrow Zu) = 0.1\%$ is shown by way of illustration. }
\label{fig:fit}
\end{figure}

The signal is extracted by using a likelihood fit on the distributions from the signal and the background enriched regions, and the four different final states, simultaneously. A multivariate analysis using boosted decision trees (BDT) is implemented for both searches.  For the tZq-SM search, the BDT is constructed in order to discriminate between the signal and the dominating ttZ and WZ backgrounds, while the tZ-FCNC searches have a BDT discriminating against all SM backgrounds. The normalisation of the non-prompt lepton and WZ background is estimated by fitting the \mT\ distribution in the background enriched region simultaneously with the BDT distribution in the signal regions (see Figure~\ref{fig:fit}).

\section{Results}
\begin{figure}[htb]
\centering
\includegraphics[height=2in]{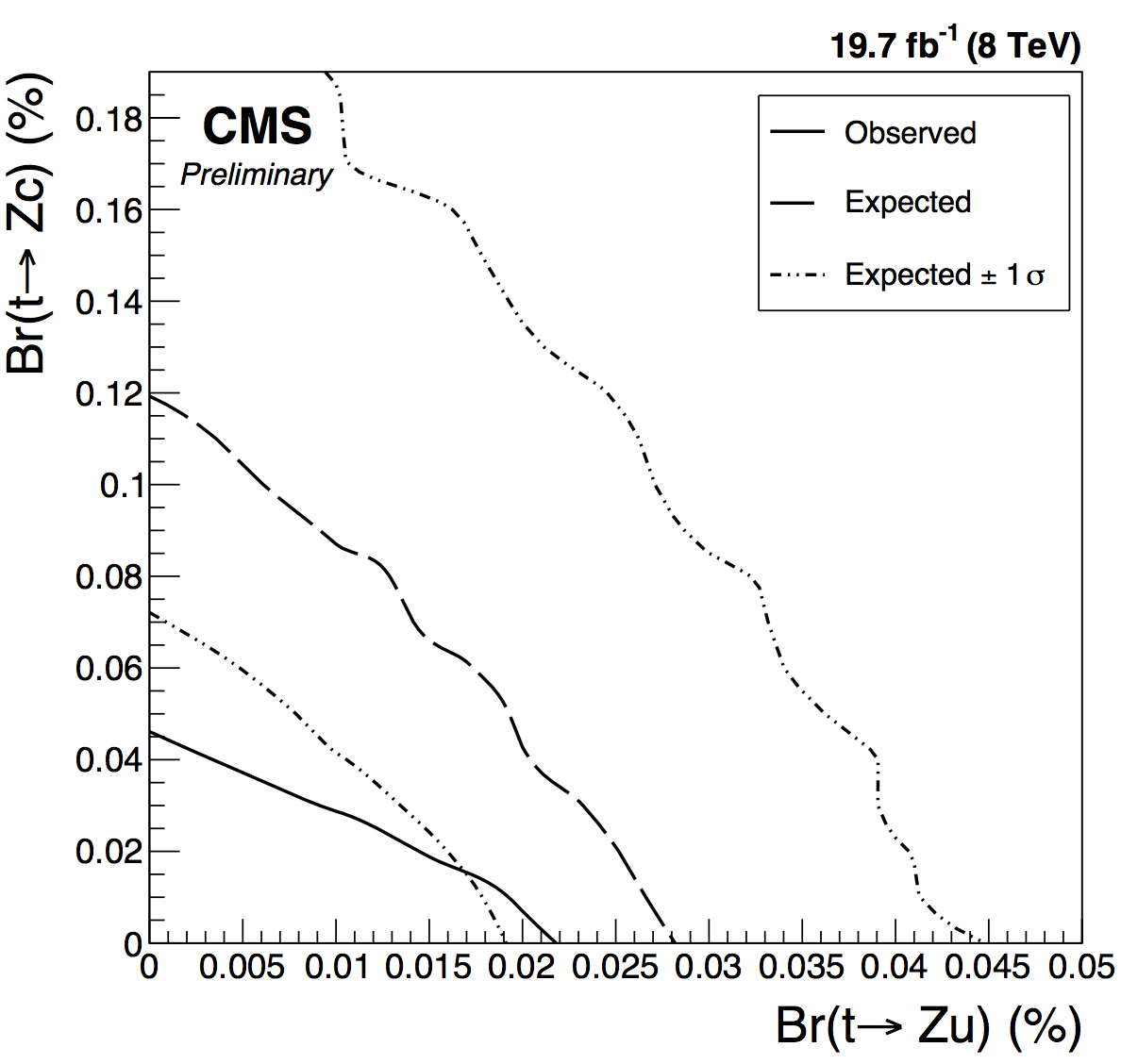}
\caption{Exclusion limits with the $\pm 1$ sigma band on the $BR(t\rightarrow Zc)$ at 95$\%$ CL  as a function of the limits on the $BR(t\rightarrow Zu)$. }
\label{fig:limit}
\end{figure}

 The combined measured signal cross section of the tZq-SM process is found to be $\sigma(tZq) = 10^{+8}_{-7}$ fb and is in agreement 
with the SM prediction of $ 8.2^{+0.59}_{-0.03}$ (scale) fb. The corresponding observed and expected significances are 2.4 and 1.8 respectively. The observed signal exclusion limit on the tZq cross section is 21 fb at the 95$\%$ CL.\\
For the tZ-FCNC search, no excess over the SM prediction is observed and exclusion limits are extracted. for different combinations of tZu and tZc anomalous couplings, as shown in Figure~\ref{fig:limit}. Exclusion limits can also be calculated independently by assuming that the BR of the coupling that is not considered is 0$\%$.  A  $95 \%$ CL   limit of $0.022\%$ is observed on the tZu coupling (expected $0.027\%$) compared to $0.049\%$ for the tZc (expected $0.118\%$).  

\end{document}